\documentclass[conference, 10pt]{IEEEtran}
\IEEEoverridecommandlockouts
\usepackage{cite}
\usepackage{bm}
\usepackage{multicol}
\usepackage{multirow}
\usepackage[cmex10]{amsmath}
\usepackage{graphicx}
\usepackage{url}
\usepackage{color}
\usepackage{amsfonts}
\usepackage{amsmath}
\usepackage{extarrows}
\usepackage{graphicx}
\usepackage{amsfonts}
\usepackage{caption}
\usepackage{algorithm}
\usepackage{algpseudocode}
\usepackage{indentfirst}
\usepackage{caption}
\usepackage{esint} 
\usepackage{graphics}
\usepackage{epsfig}
\usepackage{soul}
\usepackage{caption}
\usepackage{gensymb}
\usepackage{algpseudocode}  
\usepackage{amsmath} 
\usepackage{algorithm,algpseudocode}
\makeatletter
\newcommand\fs@spaceruled{\def\@fs@cfont{\bfseries}\let\@fs@capt\floatc@ruled
  \def\@fs@pre{\vspace{1\baselineskip}\hrule height.8pt depth0pt \kern2pt}%
  \def\@fs@post{\kern2pt\hrule\relax}%
  \def\@fs@mid{\kern2pt\hrule\kern2pt}%
  \let\@fs@iftopcapt\iftrue}
\makeatother

\def\bA{{\mathbf{A}}}   \def\bD{{\mathbf{D}}} 
 \def\bG{{\mathbf{G}}} \def\bH{{\mathbf{H}}}  
   \def\bN{{\mathbf{N}}} 
    
  \def\bW{{\mathbf{W}}} \def\bX{{\mathbf{X}}} \def\bY{{\mathbf{Y}}}

\def\ba{{\mathbf{a}}} \def\bb{{\mathbf{b}}}   
\def\bf{{\mathbf{f}}} \def\bg{{\mathbf{g}}} \def\bh{{\mathbf{h}}}  
   \def\bn{{\mathbf{n}}} 
    
  \def\bw{{\mathbf{w}}} \def\bx{{\mathbf{x}}} \def\by{{\mathbf{y}}}

\begin{document}
\title{Channel Estimation for RIS-Aided mmWave MIMO Channels
\thanks{This work is supported by Horizon 2020, European Union's Framework Programme for Research and Innovation, under grant agreement no. 871464 (ARIADNE). This work is also partially supported by the Academy of Finland 6Genesis Flagship (grant 318927) and Swedish Research Council (grant no. 2018-03701). The work of M. Leinonen has been financially supported in part by Infotech Oulu and the Academy of Finland (grant 319485 and 323698).}
}
\author{Jiguang~He$^\dag$, Markus~Leinonen$^\dag$, Henk~Wymeersch$^\star$, Markku~Juntti$^\dag$\\
        $^\dag$Centre for Wireless Communications, FI-90014, University of Oulu, Finland\\
        $^\star$Department of Electrical Engineering, Chalmers University of Technology, Gothenburg, Sweden\\}

\maketitle
\begin{abstract}
A reconfigurable intelligent surface (RIS) can shape the radio propagation by passively changing the directions of impinging electromagnetic waves. The optimal control of the RIS requires perfect channel state information (CSI) of all the links connecting the base station (BS) and the mobile station (MS) via the RIS. Thereby the channel (parameter) estimation at the BS/MS and the related message feedback mechanism are needed. In this paper, we adopt a two-stage channel estimation scheme for the RIS-aided millimeter wave (mmWave) MIMO channels using an iterative reweighted method to sequentially estimate the channel parameters. We evaluate the average spectrum efficiency (SE) and the RIS beamforming gain of the proposed scheme and demonstrate that it achieves high-resolution estimation with the average SE comparable to that with perfect CSI.

\end{abstract}

\section{Introduction}


Large unused spectrum is available in the millimeter wave (mmWave) bands. 
In order to compensate for the high free space path loss, large antenna arrays are needed  both at the transmitters and receivers~\cite{Alkhateeb2014,He2014,Heath2016}. This makes the channel estimation (CE) more difficult than that in conventional sub-6 GHz multiple-input multiple-output (MIMO) systems having less transmit and receive antennas. The mmWave MIMO channel is inherently sparse due to the limited number of distinguishable paths in the angular domain. Thus, compressive sensing (CS) techniques, which take advantage of the sparsity, have been widely applied in the channel (parameter) estimation of mmWave MIMO channels, e.g., in~\cite{Marzi2016,Lee2016}. 


In order to further improve the spectrum efficiency (SE) and to guarantee a wide communication coverage, the concept of a reconfigurable intelligent surface (RIS) has been recently proposed to smartly shape the propagation of electromagnetic waves~\cite{Hu2018_SP,Huang2018,Hu2018,He2019large}. The RIS also has great potential to offer higher accuracy of positioning and localization, both for indoor and outdoor, compared to the system without RISs~\cite{He2019large,wymeersch2019radio}. The RIS can be made of an array of phase shifters, which can passively steer the beams towards the dedicated user(s) by controlling the phase of each RIS unit. This RIS architecture is called the discrete RIS. Another type of RIS 
is the contiguous RIS which can be seen as an active transceiver~\cite{Hu2018_SP}. 

CE for RIS-aided MIMO systems has been recently studied in~\cite{Taha2019,He2020}. In~\cite{Taha2019}, CE is performed using CS and deep learning methods
in a setup with a few active elements at the RIS. In~\cite{He2020}, sparse matrix factorization and matrix completion are exploited in a sequential manner to perform iterative CE. In this work, full advantage of the RIS is not achieved due to the on/off state applied to the RIS elements. Instead of estimating the MIMO channels, a multi-level codebook based scheme was leveraged to design the phase control matrix at the RIS and the combining vector at the MS jointly~\cite{he2019adaptive}.

In the paper, we study the CE problem of the passive RIS-aided mmWave MIMO system. We divide the CE problem into two subproblems and apply
an iterative reweighted method to find the estimates of the channel parameters sequentially. In the first stage, we estimate the angle of departure (AoD) for the BS-RIS link, the angle of arrival (AoA) of the RIS-MS link, and the corresponding effective propagation path gains. In the second stage, we estimate the product of the propagation path gains of the BS-RIS link and the RIS-MS link and the difference of directional sine of the AoA of the BS-RIS and the AoD of RIS-MS link. Besides evaluating the mean square error (MSE) of the estimated channel parameters, we design the RIS phase control matrix, the BS beamforming (BF) vector, and the MS combining vector based on the estimates and evaluate the average SE and RIS BF gain. The performance of the proposed scheme is compared to that of an orthogonal matching pursuit (OMP) approach. Simulation results demonstrate that average SE achieved by the proposed method approaches that with perfect channel state information (CSI), even in the low signal-to-noise ratio (SNR) regime with limited training overhead. 

\textit{Notation:} A bold lowercase letter $\ba$ denotes the column vector, a bold capital letter $\bA$ denotes the matrix, $(\cdot)^{\mathsf{H}}$, $(\cdot)^{\mathsf{T}}$, and $(\cdot)^*$ denote the Hermitian transpose, transpose, and conjugate, respectively, $\mathrm{diag}(\ba)$ denotes a square diagonal matrix with entries of $\ba$ on its diagonal, $\ba \circ \bb$ denotes the Hadamard product of $\ba$ and $\bb$, $[\ba]_i$ denotes the $i$th element of vector $\ba$, $[\bA]_{ij}$ denotes the $(i,j)$th element of~$\bA$, and $\|\cdot\|_{\mathrm{F}}$ is the Frobenius norm.

\section{Channel Model}
\begin{figure}[t]
	\centering
	\includegraphics[width=0.75\linewidth]{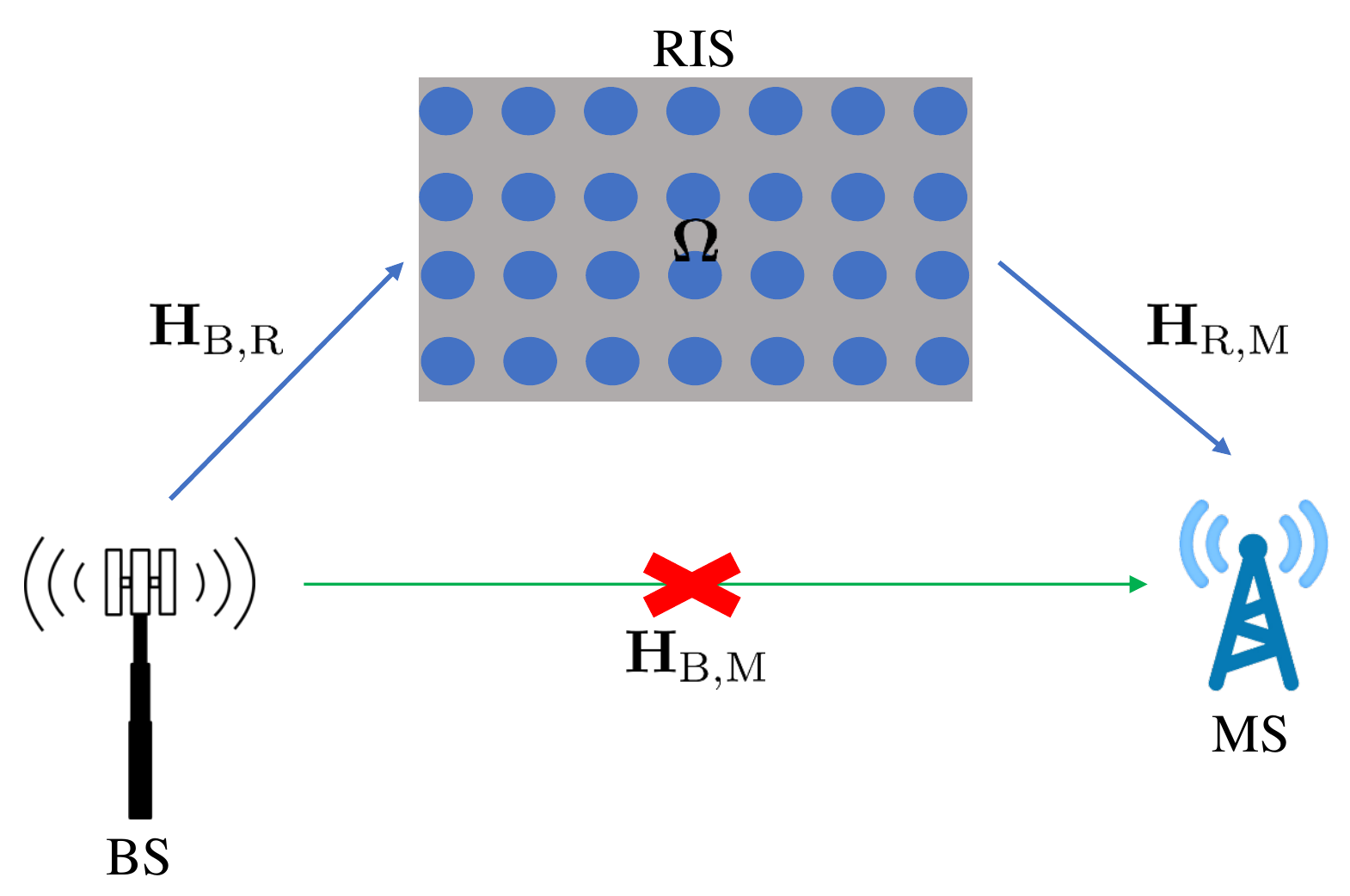}
	\caption{The considered RIS-aided mmWave MIMO system with one BS, one MS, and one RIS.}
	\label{System_model}
\end{figure}
We consider the RIS-aided mmWave MIMO system that comprises one multi-antenna BS, one multi-antenna MS, and one multi-element RIS, as depicted in Fig.~\ref{System_model}. The numbers of antenna elements are defined as $N_{\text{BS}}$, $N_{\text{MS}}$, and $N_{\text{RIS}}$, respectively. The antenna array is assumed to be an uniform linear array (ULA); an extension to an uniform planar array (UPA) is feasible. We assume that the direct link between the BS and the MS are obstructed, which renders the potential usage of a RIS for maintaining the connectivity. 

We assume geometric channel model. 
The channel between the BS and the RIS $\bH_{\text{B,R}} \in \mathbb{C}^{N_{\text{RIS}} \times N_{\text{BS}}}$ is 
\begin{equation}\label{BS_RIS}
\bH_{\text{B,R}} = \sum\limits_{l = 0}^{L_{\text{B,R}}} \rho_{\text{B,R},l} \boldsymbol{ \alpha}(\phi_{\text{B,R},l} ) \boldsymbol{\alpha}^{\mathsf{H}}(\theta_{\text{B,R},l} ),
\end{equation}
where $\theta_{\text{B,R},l}$ and $\phi_{\text{B,R},l}$ denote the $l$th AoD and AoA of the BS-RIS link, respectively, $L_{\text{B,R}}$ denotes the number of resolvable paths, which is usually on the order of 3--5 in mmWave bands~\cite{Rappaport2013}, and $\rho_{\text{B,R},l}$ denotes the $i$th propagation path gain. Index ${l=0}$ refers to the line-of-sight (LoS) path, and $l=1, \cdots, L_{\text{B,R}}$ refer to the non-line-of-sight (NLoS) paths, e.g., single-bounce or multi-bounce reflection paths. Usually, $|\rho_{\text{B,R},0}|^2 \gg |\rho_{\text{B,R},l}|^2$ for $l= 1,\cdots, L_{\text{B,R}}$, and the difference is easily more than 20 dB~\cite{Akdeniz2014}. Finally, ${\boldsymbol{ \alpha}(\theta_{\text{B,R},l} ) \in \mathbb{C}^{N_{\text{BS}} \times 1}}$ and ${\boldsymbol{\alpha}(\phi_{\text{B,R},l})\in \mathbb{C}^{N_{\text{RIS}} \times 1}}$ are the array response vectors with $[\boldsymbol{ \alpha}(\theta_{\text{B,R},l} )]_k = \exp\big(j 2 \pi \frac{d}{\lambda} (k-1) \sin(\theta_{\text{B,R},l})\big)$ for $k =1,\cdots, N_{\text{BS}}$ and $[\boldsymbol{\alpha}(\phi_{\text{B,R},l})]_k = \exp\big(j 2 \pi \frac{d}{\lambda} (k-1) \sin(\phi_{\text{B,R},l})\big)$ for $k =1,\cdots, N_{\text{RIS}}$, where $d$ is the antenna element spacing, $\lambda$ is the wavelength of the carrier frequency, and $j \overset{\triangle}{=} \sqrt{-1}$. 

Similarly, the channel between the RIS and the MS, denoted as $\bH_{\text{R,M}} \in \mathbb{C}^{N_{\text{MS}} \times N_{\text{RIS}}}$, is 
\begin{equation}\label{RIS_MS}
\bH_{\text{R,M}} = \sum\limits_{l = 0}^{L_{\text{R,M}}} \rho_{\text{R,M},l} \boldsymbol {\alpha}(\phi_{\text{R,M},l} ) \boldsymbol{\alpha}^{\mathsf{H}}(\theta_{\text{R,M},l} ). 
\end{equation}

Using~\eqref{BS_RIS} and~\eqref{RIS_MS}, the composite channel ${\bH \in \mathbb{C}^{N_{\text{MS}} \times N_{\text{BS}}}}$ between the BS and MS becomes 
\begin{equation}\label{Entire_channel}
\bH =  \bH_{\text{R,M}} \boldsymbol{\Omega} \bH_{\text{B,R}},
\end{equation}
where $\boldsymbol{\Omega}\in\mathbb{C}^{N_{\text{RIS}}\times N_{\text{RIS}}}$ is the phase control matrix at the RIS. The matrix $\boldsymbol{\Omega}$ is a diagonal matrix with unit-modulus elements on the diagonal, i.e., $[\boldsymbol{\Omega}]_{kk} = \exp(j \omega)$ with $\omega \in \mathbb{R}$. In practice, the reflection may not be perfect so that reflection coefficient $a \in [0,\;1]$ as $[\boldsymbol{\Omega}]_{kk} = a \exp(j \omega)$ describes the amplitude scaling and power loss \cite{Wu2019}; however, we assume that $a = 1$.  

Since the LoS path is typically much stronger than the NLoS paths according to the field measurements in~\cite{Akdeniz2014}, we ignore the NLoS paths in the BS-RIS and RIS-MS links, and approximate the composite channel in~\eqref{Entire_channel}  as 
\begin{align}
\bH & \approx \boldsymbol {\alpha}(\phi_{\text{R,M},0}) \rho_{\text{R,M},0} \boldsymbol {\alpha}^{\mathsf{H}}(\theta_{\text{R,M},0}) \boldsymbol{\Omega} 
\boldsymbol {\alpha}(\phi_{\text{B,R},0}) \rho_{\text{B,R},0} \boldsymbol {\alpha}^{\mathsf{H}}(\theta_{\text{B,R},0}) \nonumber\\ \label{Simplified_channel}
& = g \boldsymbol {\alpha}(\phi_{\text{R,M},0})  \boldsymbol {\alpha}^{\mathsf{H}}(\theta_{\text{B,R},0}), 
\end{align} 
where $g \in \mathbb{C}$ is 
the effective propagation path gain, defined as
\begin{equation}\label{G_matrix}
g =  \rho_{\text{B,R},0} \rho_{\text{R,M},0} \boldsymbol {\alpha}^{\mathsf{H}}(\theta_{\text{R,M},0}) \boldsymbol{\Omega} \boldsymbol {\alpha}(\phi_{\text{B,R},0}).
\end{equation}
According to~\eqref{Simplified_channel} and~\eqref{G_matrix}, the composite channel is virtually a point-to-point MIMO channel with one path via the RIS. The rank of $\bH$ is $1 \ll \{N_\text{BS}, N_\text{MS}\}$. Thus, the channel encompasses a sparse structure in the angular domain -- characteristic to an mmWave channel -- which can efficiently be leveraged by CS. 

\section{Sounding Procedure}
\begin{figure}[t]
	\centering
	\includegraphics[width=0.7\linewidth]{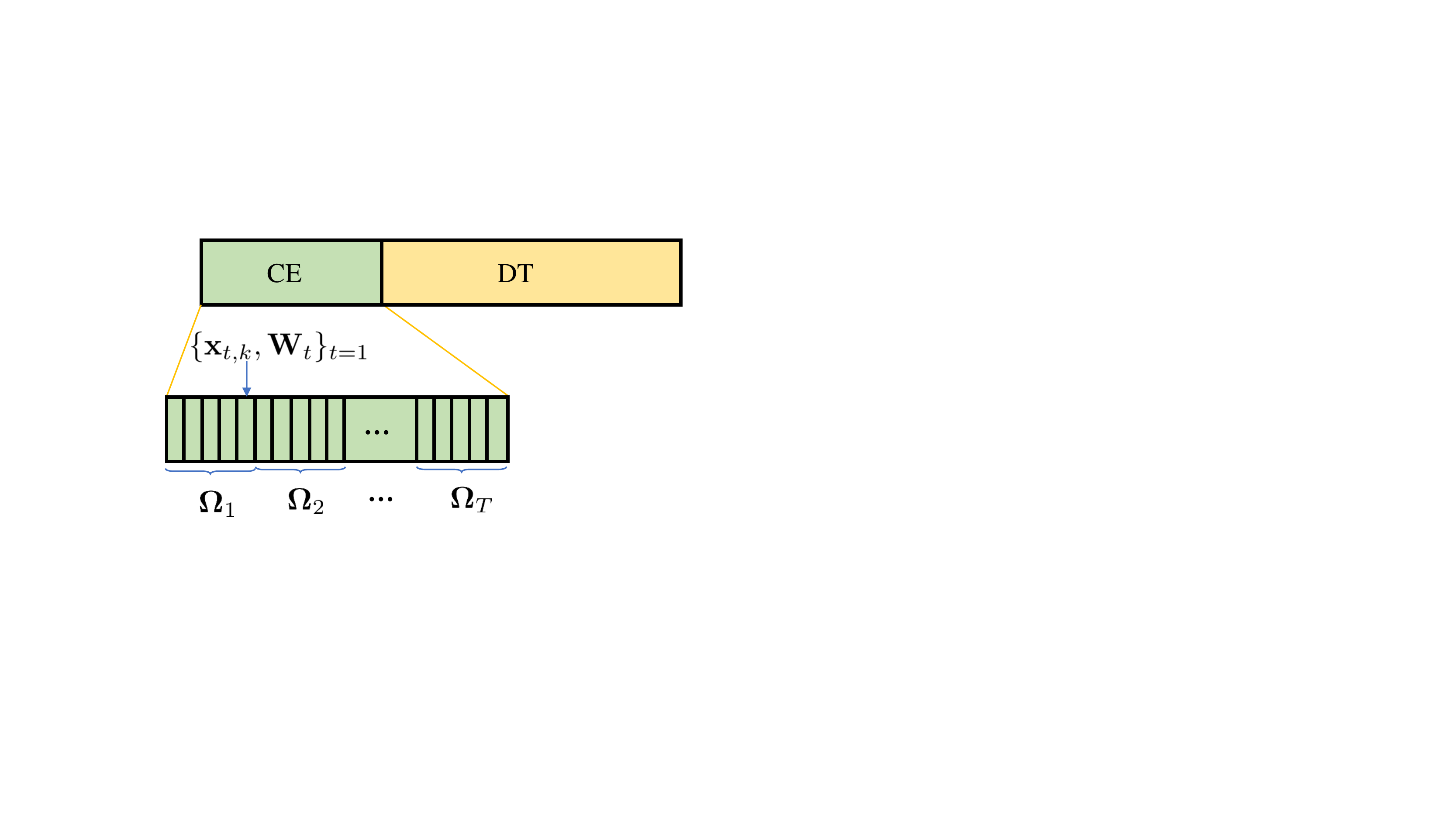}
	\caption{The sounding and CE procedure. Each CE subinterval contains multiple blocks where $\boldsymbol{\Omega}_t$ varies over the blocks.}
	\label{Sounding_procedure}
\end{figure}

We assume that the channels suffer from block fading. For the sounding process, one coherence time interval is divided into two subintervals, the first one for CE and the second for data transmission (DT), as depicted in Fig.~\ref{Sounding_procedure}. The CE subinterval is further divided into multiple blocks.

In each block ${t=1,\cdots,T}$, the BS sends a (random) training matrix $\bX_t \in \mathbb{C}^{N_{\text{BS}}\times N_X}$ which, after reflected from the RIS having a (random) phase control matrix $\boldsymbol{\Omega}_t$, is received at the MS through a (random) combining matrix $\bW_t \in \mathbb{C}^{N_{\text{MS}}\times N_W}$. Thus, the received signal at the MS is given as 
\begin{equation}
\bY_t = \bW_t^{\mathsf{H}} \bH(\boldsymbol{\Omega}_t) \bX_t + \bW_t^{\mathsf{H}} \bN_t,\;t = 1, \cdots, T,
\end{equation}
where we write $\bH$ explicitly as a function of $\boldsymbol{\Omega}_t$. Further, let 
\begin{equation}\label{G_matrix_training}
g_t = \rho_{\text{R,M},0} \rho_{\text{B,R},0} \boldsymbol{\alpha}^{\mathsf{H}}(\theta_{\text{R,M},0}) \boldsymbol{\Omega}_t \boldsymbol{\alpha}(\phi_{\text{B,R},0}).
\end{equation}
The following assumptions are made for the CE subinterval: 
\begin{itemize}
    \item All the channel parameters stay unchanged within the coherence time. 
    \item $\boldsymbol{\Omega}_t$ is varying with block index $t$, and it is constructed with randomly generated phases $\omega\in[-\pi \; \pi]$.
    \item $\bW_t$ and $\bX_t$ are varying with block index $t$.
\end{itemize}

\section{Two-Stage CE Approach}\label{CE_Approach}
We consider a two-stage approach to simplify the CE problem. The solution process is summarized in Algorithm~\ref{CE_Algorithm}. The details of the algorithm will be provided in the sequel.

\floatstyle{spaceruled}
\restylefloat{algorithm}
    \begin{algorithm}[H]
      \caption{Two-stage CE Approach}\label{CE_Algorithm}  
      \begin{algorithmic}[1]  
      \Statex $\triangleright$ \textit{First stage}
        \Require  
          Received signals $\bY_1,\dots, \bY_T$, combining matrices
          $\bW_1,\dots, \bW_T$, training matrices $\bX_1,\dots, \bX_T$, phase control matrices
          $\boldsymbol{\Omega}_1,\dots, \boldsymbol{\Omega}_T$, and threshold value $\tau_{\text{th}}$.
        \Ensure  
          $\hat{\theta}, \hat{\phi}$, and $\hat{\bg}$.
          \State Initial $\theta^{(i)}$ and $\phi^{(i)}$ for $i = 0$ by the SVD based approach. 
                \Repeat  
          \State Compute $\hat{g}_t^{(i+1)}$ by~\eqref{Optimal_g}.  
          \State Construct the objective function $\tilde{S}(\hat{\bg}^{(i+1)},\theta, \phi)$ by~\eqref{obj_func1}.  
          \State Use gradient descent to minimize $\tilde{S}(\hat{\bg}^{(i+1)},\theta, \phi)$.   
        \Until{The maximum number of iterations reached or $\|\hat{\bg}^{(i+1)} - \hat{\bg}^{(i)}\|_2 < \tau_{\text{th}}$}.  
        
      \Statex $\triangleright$ 	 \textit{Second stage}
          \Require $\hat{\bg}$ and threshold value $\tilde{\tau}_{\text{th}}$.
         \Ensure  
          $ \hat{q}$ and $\hat{\tilde{\theta}}$.
          \State Initial $\tilde{\theta}^{(i)}$ for $i =0$. 
                \Repeat  
          \State Compute $\hat{q}^{(i+1)}$.
         \State Construct the objective function $\tilde{J}(\hat{q}^{(i+1)}, \tilde{\theta})$ by~\eqref{obj_second_step1}.  
          \State Use gradient descent to minimize $\tilde{J}(\hat{q}^{(i+1)}, \tilde{\theta})$.   
        \Until{The maximum number of iterations reached or $|\hat{q}^{(i+1)} - \hat{q}^{(i)}| < \tilde{\tau}_{\text{th}}$}.  
      \end{algorithmic}  
    \end{algorithm}

\subsection{First Stage}
In the first stage, we estimate the AoD of the BS-RIS link $\theta_{\text{B,R},0}$ in~\eqref{Simplified_channel}, the AoA of the RIS-MS link $\phi_{\text{R,M},0}$ in~\eqref{Simplified_channel}, and the effective propagation path gain of the reflection link $g_t$ in \eqref{G_matrix_training} based on the received signals $\{\bY_t\}_{t=1}^T$.  
Typically, the CS methods like OMP have been proposed for CE. One drawback is that they only can recover angular estimates that lie on a pre-defined grid of discrete angles. This ``on-the-grid'' problem, which inevitably deteriorates the CE performance, can be counteracted by the iterative reweighted method~\cite{Hu2018}, used herein as well. 

Using the iterative reweighted method, the CE problem in the first stage is formulated as 
\begin{equation}\label{Fisrt_step_obj}
    \underset{\bg, \theta, \phi}{\min} S(\bg, \theta, \phi) = \sum_{t = 1}^T \ln(|g_t|^2 + \epsilon) + \xi \|\bY_t - \bW_t^{\mathsf{H}} \bH(\boldsymbol{\Omega}_t)\bX_t\|^2_{\mathrm{F}},
\end{equation}
where $\bg = [g_1,\cdots, g_T]^{\mathsf{T}}$, parameter $\epsilon > 0$ ensures that the argument of $\mathrm{ln}(\cdot)$ is positive, $\xi > 0$ controls the tradeoff between the sparsity of $\bg$ and data fitting. Here, we replaced $\phi_{\text{R,M},0}$ and $\theta_{\text{B,R},0}$ with $\phi$ and $\theta$ to simplify the notations. The sparsity-inducing $\mathrm{ln}(\cdot)$ term accounts for the fact that due to the random generation of $\boldsymbol{\Omega}_t$, $t=1,\cdots,T$, some elements in $\bg$ may be much smaller than others. 

The problem in~\eqref{Fisrt_step_obj} can be further formulated as~\cite{Fang2016}
\begin{equation}\label{obj_func}
    \min_{\bg, \theta, \phi} \tilde{S}(\bg, \theta, \phi) = \bg^{\mathsf{H}}\bG^{(i)}\bg +\xi \sum_{t = 1}^T \|\bY_t - \bW_t^{\mathsf{H}} \bH(\boldsymbol{\Omega}_t)\bX_t\|^2_{\mathrm{F}}, 
\end{equation}
where 
\begin{equation}
    {\bG^{(i)} = \mathrm{diag}\Big(\big[\frac{1}{|\hat{g}^{(i)}_1|^2 + \epsilon},\cdots,\frac{1}{|\hat{g}^{(i)}_T|^2 + \epsilon}\big]^{\mathsf{T}}\Big)} \nonumber
\end{equation}
and $\hat{g}^{(i)}_t$ is an estimate of $g_t$ at the $i$th iteration. Setting the first-order partial derivative of $\tilde{S}(\bg, \theta, \phi)$ in \eqref{obj_func} with respect to $g_t$ to zero
yields 
\begin{equation}\label{Optimal_g}
    \hat{g}_t^{(i+1)} =  \Bigg(\frac{1}{\xi (|\hat{g}^{(i)}_t|^2 + \epsilon) } + \|\bA_t\bX_t\|_{\mathrm{F}}^2\Bigg)^{-1} \sum_{k = 1}^{N_X} \bx_{t,k}^{\mathsf{H}}\bA_t^{\mathsf{H}}\by_{t,k}, 
\end{equation}
where $\bx_{t,k}$ and $\by_{t,k}$ is the $k$th column of $\bX_t$ and $\bY_t$, respectively, and $\bA_t = \bW_t^{\mathsf{H}} \boldsymbol {\alpha}(\phi) \boldsymbol {\alpha}^{\mathsf{H}}(\theta)$. Thus, for given $\hat{\bg}^{(i+1)}= [\hat{g}_1^{(i+1)},\cdots, \hat{g}_T^{(i+1)}]^{\mathsf{T}}$, \eqref{obj_func} can be written as 
\begin{align}\label{obj_func1}
   & \tilde{S}(\hat{\bg}^{(i+1)},\theta, \phi) =\frac{1}{\xi} \sum_{t =1}^T \beta_t^2 z_t^* z_t \frac{1}{|\hat{g}^{(i)}_t|^2 + \epsilon} +\sum_{t =1}^T\sum_{k =1}^{N_X} \by_{t,k}^{\mathsf{H}} \by_{t,k}\nonumber\\
    &+\sum_{t =1}^T \left\{-\hat{g}_t^{(i+1)} z_t^*- \left(\hat{g}_t^{(i+1)}\right)^* z_t + |\hat{g}_t^{(i+1)}|^2 \|\bA_t\bX_t\|_{\mathrm{F}}^2 \right\}\nonumber\\
    &= \sum_{t =1}^T \left(- \beta_t z_t^* z_t + \sum_{k =1}^{N_X} \by_{t,k}^{\mathsf{H}} \by_{t,k}\right),
\end{align}
where we defined the quantities ${z_t = \sum\limits_{k = 1}^{N_X} \bx_{t,k}^{\mathsf{H}}\bA_t^{\mathsf{H}}\by_{t,k}}$ and ${\beta_t = \Big(\frac{1}{\xi (|\hat{g}^{(i)}_t|^2 + \epsilon) } + \|\bA_t\bX_t\|_{\mathrm{F}}^2\Big)^{-1}}$. 

The above equations give rise to an iterative algorithm aiming at minimizing the objective function in~\eqref{obj_func1}. To this end, we propose Algorithm~\ref{CE_Algorithm} where at each iteration $i$, we use a gradient descent algorithm to find estimates for the angles $\theta$ and $\phi$ for a given $\hat{\bg}^{(i+1)}$ in~\eqref{Optimal_g}.
The calculation of the required first-order partial derivatives associated with $\theta$ and $\phi$ is presented in Appendix~\ref{FirstAppendix}.  
The initial values for $\theta$ and $\phi$, defined as $\theta^{(0)}$ and $\phi^{(0)}$, can be determined by the singular value decomposition (SVD) based approach~\cite{Hu2018}.
After a certain stopping criterion is met, we proceed to the second stage.

\subsection{Second Stage}
In the second stage, we estimate the remaining channel parameters based on the final estimate of $\bg$ obtained in the first stage, denoted as $\hat{\bg} = [\hat{g}_1, \cdots, \hat{g}_T]^{\mathsf{T}}$. 
Obtaining a separate estimate of the AoA for the BS-RIS link $\phi_{\text{B,R},0}$ and the AoD for the RIS-MS link $\theta_{\text{R,M},0}$ seems infeasible; the same holds for the propagation path gains in the BS-RIS and RIS-MS links $\rho_{\text{R,M},0}$ and $\rho_{\text{B,R},0}$. Thus, instead, we will estimate the product of the propagation path gains $ \rho_{\text{R,M},0} \rho_{\text{B,R},0}$ and the difference of directional sine, defined as $\theta_{\text{diff}} = \sin(\phi_{\text{R,M},0})-\sin(\theta_{\text{B,R},0})$. 

According to~\eqref{G_matrix_training}, we rewrite $g_t$ as 
\begin{equation}
    g_t = \rho_{\text{R,M},0} \rho_{\text{B,R},0} \boldsymbol{\omega}_t^{\mathsf{T}} \big(\boldsymbol {\alpha}^*(\theta_{\text{R,M},0}) \circ \boldsymbol {\alpha}(\phi_{\text{B,R},0})\big),
\end{equation}
where ${\boldsymbol{\omega}_t\in\mathbb{C}^{N_{\text{RIS}}\times{1}}}$ is the vector of phase control matrix values as ${\boldsymbol{\Omega}_t = \mathrm{diag}(\boldsymbol{\omega}_t)}$. By stacking the $T$ different $\boldsymbol{\omega}_t$'s row-wise as $\tilde{\boldsymbol{\Omega}} = [\boldsymbol{\omega}_1, \cdots, \boldsymbol{\omega}_T ]^{\mathsf{T}}$ and introducing $q =  \rho_{\text{R,M},0} \rho_{\text{B,R},0}$ and $\tilde{\theta} = \mathrm{asin} (\theta_{\text{diff}})$, we have 
\begin{equation}
    \bg \!= \tilde{\boldsymbol{\Omega}} \rho_{\text{R,M},0} \rho_{\text{B,R},0}  \big(\boldsymbol {\alpha}^*(\theta_{\text{R,M},0}) \!\circ\! \boldsymbol {\alpha}(\phi_{\text{B,R},0})\big) \!= q \tilde{\boldsymbol{\Omega}}  \boldsymbol {\alpha}(\tilde{\theta}).
\end{equation}

Recall that the the angular sparsity is one. Regarding this fact, we aim at finding the sparsest representation while minimizing the MSE of the data fitting, which fits in the problem of line spectral estimation~\cite{Fang2016}. Further, we assume that 
\begin{equation}
    \hat{\bg} =\bg+\bn = q \tilde{\boldsymbol{\Omega}} \boldsymbol {\alpha}(\tilde{\theta})+\bn,
\end{equation}
where the estimation error $\bn$ from the first stage is modelled as additive Gaussian noise, independent of $q$ and $\tilde{\theta}$. Thus, the objective function of the second stage estimation problem is formulated as 
\begin{equation}
    \min J(q, \tilde{\theta}) =  \ln(|q|^2 + \epsilon) + \mu \|\hat{\bg} - \tilde{\boldsymbol{\Omega}} \bh\|_2^2,
\end{equation}
where $\bh = q  \boldsymbol {\alpha}(\tilde{\theta})$ and $\mu$ is a parameter that controls the tradeoff between the sparsity and data fitting. The objective function can be further reformulated as
\begin{equation}\label{obj_second_step}
   \min \tilde{J}(q, \tilde{\theta}) =  \frac{q q^*}{|\hat{q}^{(i)}|^2 + \epsilon} +\mu \|\hat{\bg} - \tilde{\boldsymbol{\Omega}} \bh\|_2^2.
\end{equation}

We also follow the iterative reweighted method to find the high-resolution estimate of $\tilde{\theta}$. At the beginning, we find an initial estimate of $\tilde{\theta}$ on the grid as
    $\tilde{\theta}^{(0)} = \arg\max_{\tilde{\theta} \in \Upsilon} |\hat{\bg}^{\mathsf{H}} \tilde{\boldsymbol{\Omega}} \bh|,$
where $\Upsilon$ is a quantized set of angles within $[-\pi\; \pi]$. 
Setting the derivative of the objective function $\tilde{J}(q, \tilde{\theta})$ in \eqref{obj_second_step} with respect to $q$ to zero
yields $\hat{q}^{(i+1)}= \gamma \boldsymbol {\alpha}(\tilde{\theta})^{\mathsf{H}}\tilde{\boldsymbol{\Omega}}^{\mathsf{H}}\hat{\bg}$, 
where $\gamma = \big({1}/({\mu(|\hat{q}^{(i)}|^2 + \epsilon)}) + \boldsymbol {\alpha}(\tilde{\theta})^{\mathsf{H}} \tilde{\boldsymbol{\Omega}}^{\mathsf{H}}\tilde{\boldsymbol{\Omega}}\boldsymbol {\alpha}(\tilde{\theta})\big)^{-1}$. 
For fixed $\hat{q}^{(i+1)}$, the objective function can be further written as 
\begin{equation}\label{obj_second_step1}
  \tilde{J}(\hat{q}^{(i+1)}, \tilde{\theta})=   - \gamma  \tau  + \hat{\bg}^{\mathsf{H}} \hat{\bg}.
\end{equation}
where $\tau = \boldsymbol {\alpha}(\tilde{\theta})^{\mathsf{H}}\tilde{\boldsymbol{\Omega}}^{\mathsf{H}}\hat{\bg}\hat{\bg}^{\mathsf{H}} \tilde{\boldsymbol{\Omega}}\boldsymbol {\alpha}(\tilde{\theta})$. Based on the first-order derivative in Appendix~\ref{SecondAppendix}, gradient descent is applied to find a (suboptimal) $\tilde{\theta}$ that minimizes the objective function in~\eqref{obj_second_step1}.

\section{Performance Evaluation}
In this section, we evaluate the MSE performance of the angular parameter estimation in the first stage, the average SE, and the RIS BF gain based on the estimated parameters. 

The simulation parameters are set as follows: $N_{\text{BS}} = N_{\text{BS}} = 32$, $N_{\text{RIS}} = 64$, $N_X = N_W =  10$, $T = \{10,16\}$, $\epsilon =1$, $\xi = \mu = 1000$, and $d = \frac{\lambda}{2}$. The number of RF chains is defined as $N_\text{RF} = 10$. The number of required time slots for CE is $T \frac{ N_X N_W}{N_\text{RF}} = \{100, 256\}$. We assume that the product of the propagation path gains, $q$, follows $\mathcal{CN}(0,1)$ and each element of $\bN_t$ follows $\mathcal{CN}(0,\sigma^2)$. The SNR is defined as $\frac{1}{\sigma^2}$, and 2000 realizations are considered for averaging. 

The average SE is defined as
\begin{equation}
    R = E\big[\log_2(1 +  |\bw^{\mathsf{H}} \bH(\boldsymbol{\hat{\Omega}}) \bf|^2/\sigma^2)\big] \;\; \text{bits/s/Hz}, 
\end{equation}
where $\bw = \sqrt{1/N_{\text{MS}}} \boldsymbol {\alpha}(\hat{\phi}_{\text{R,M},0})$, $\bf = \sqrt{1/N_{\text{BS}}} \boldsymbol {\alpha}(\hat{\theta}_{\text{B,R},0})$, and the estimate based optimal phase control matrix is $[\boldsymbol{\hat{\Omega}}]_{kk} = \exp(-j 2 \pi \frac{d}{\lambda}(k-1) \sin(\hat{\tilde{\theta}}) )$. The RIS BF gain is defined as 
    $G_{\text{BF}} = | \boldsymbol {\alpha}^{\mathsf{H}}(\theta_{\text{R,M},0}) \boldsymbol{\hat{\Omega}} \boldsymbol {\alpha}(\phi_{\text{B,R},0}) |/N_{\text{RIS}}.$

As a benchmark, we consider a two-stage OMP algorithm that has the following one-to-one correspondence with our proposed scheme: the first stage applies the simultaneous OMP whereas the standard OMP is uses in the second stage. The dictionary for the first stage is constructed by quantizing the angles as ${\boldsymbol{\Psi}_1 = \big(\bA^*(\{\bar{\theta}_k\}_{k =1}^{N_{\text{BS,q}}}) \otimes \bA(\{\bar{\phi}_k\}_{k =1}^{N_{\text{MS,q}}})\big)}$ with ${N_{\text{BS,q}} = 2 N_{\text{BS}}}$ and ${N_{\text{MS,q}} = 2 N_{\text{MS}}}$; the dictionary in the second stage is ${\boldsymbol{\Psi}_2 = \bA\big(\{\bar{\theta}_k\}_{k =1}^{N_{\text{RIS,q}}}\big)}$ with ${{N_{\text{RIS,q}}} = 2 N_{\text{RIS}}}$. ${\bA\big(\{\bar{\phi}_k\}_{k = 1}^{N_{\text{MS,q}}}\big)}$ is defined as ${\bA\big(\{\bar{\phi}_k\}_{k = 1}^{N_{\text{MS,q}}}\big)=\big[\boldsymbol{\alpha}(\bar{\phi}_1),\cdots, \boldsymbol{\alpha}(\bar{\phi}_{N_{\text{MS,q}}})\big]}$, and the same principle is applied to $\bA(\{\bar{\theta}_k\}_{k =1}^{N_{\text{BS,q}}})$ and $\bA(\{\bar{\theta}_k\}_{k =1}^{N_{\text{RIS,q}}})$. 

\begin{figure}[t]
	\centering
	\includegraphics[width=.75\linewidth,trim={0 0 0 3},clip]{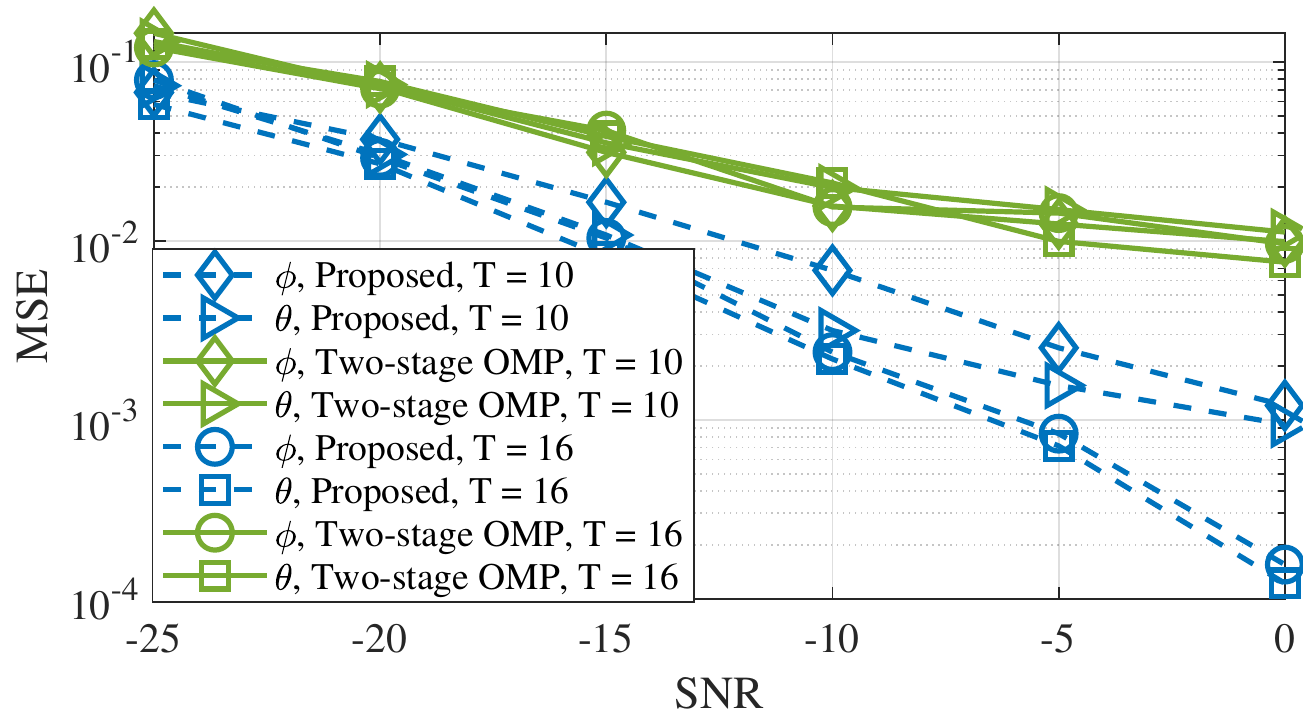}
	\caption{Angular parameter estimate in the first stage.}
	\label{Angle_estimate}
\end{figure}

\begin{figure}[t]
	\centering
	\includegraphics[width=.75\linewidth,trim={0 0 3 6},clip]{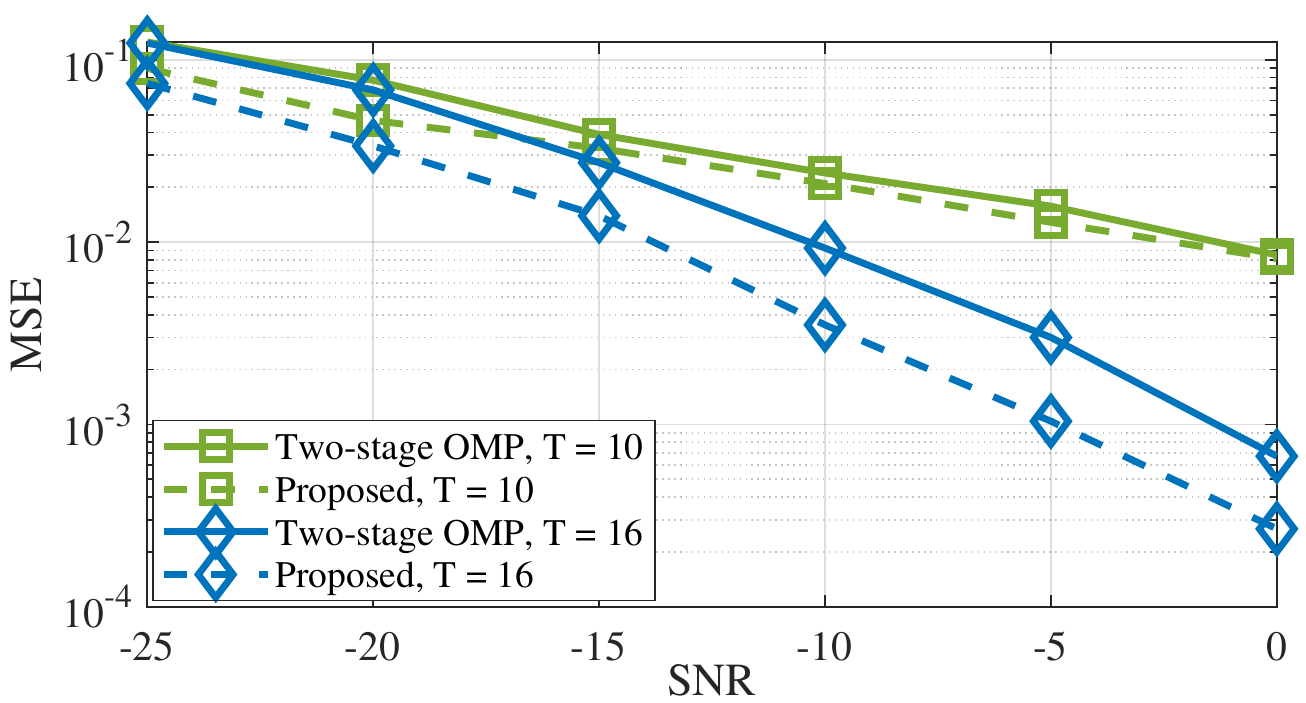}
	\caption{Angular difference estimate in the second stage.}
	\label{Angle_diff_estimate}
\end{figure}

\begin{figure}[t]
	\centering
	\includegraphics[width=.75\linewidth,trim={0 0 0 12},clip]{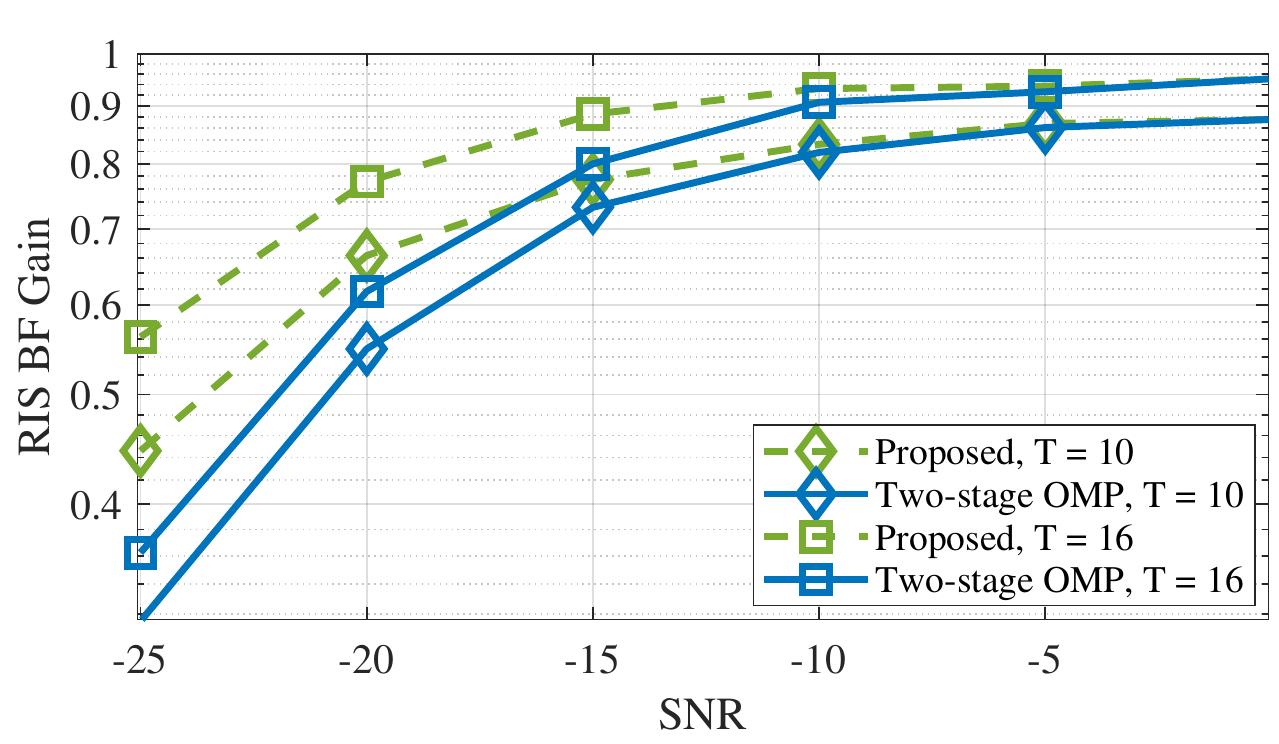}
	\caption{RIS BF gain vs. SNR.}
	\label{RIS_BF_gain}
\end{figure}

\begin{figure}[t]
	\centering
	\includegraphics[width=.75\linewidth,trim={0 0 3 6},clip]{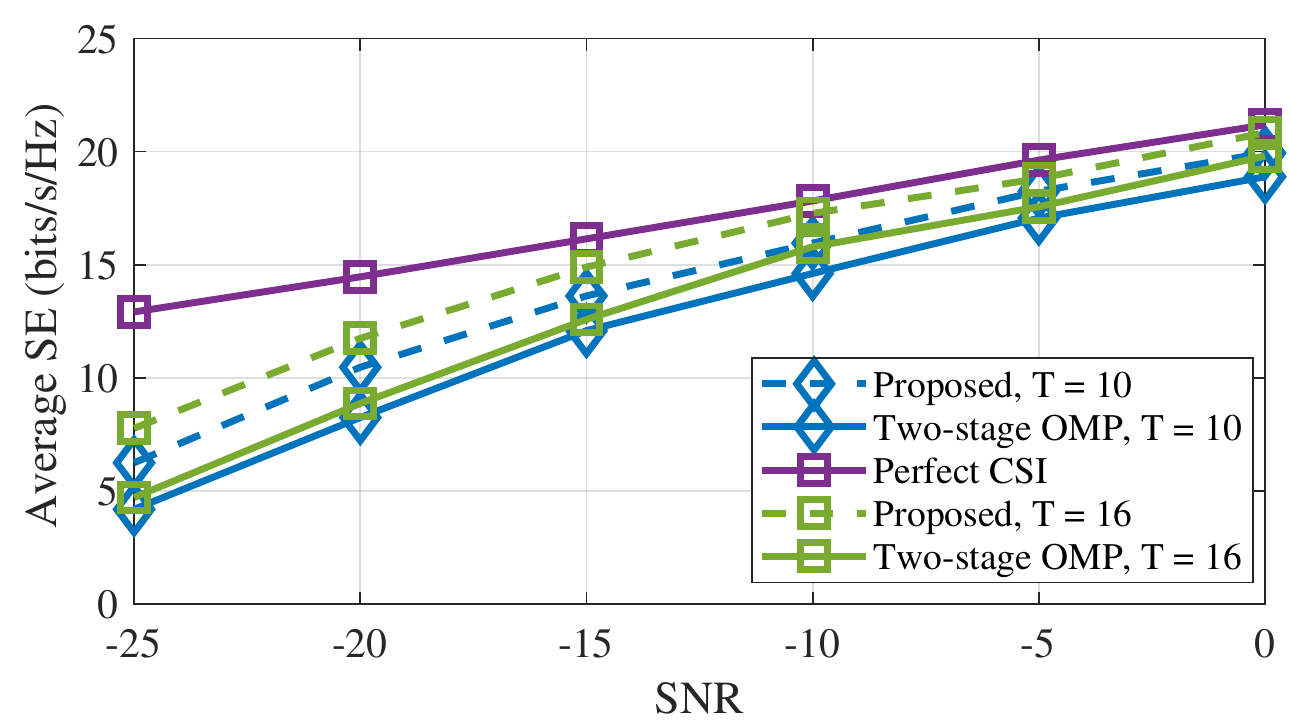}
	\caption{Average SE vs. SNR.}
	\label{SE}
\end{figure}

The simulation results are shown in Figs.~\ref{Angle_estimate}--\ref{SE}. As shown in Fig.~\ref{Angle_estimate}, the proposed scheme outperforms the two-stage OMP scheme in terms of the MSE of the AoA and AoD estimates obtained in the first stage.
A super-resolution estimate can be achieved, e.g., the average MSE is at the level of $10^{-3}$ at the SNR of $-5$ dB when $T = 16$. This offers a near-optimal design of the BF and combining vectors at the BS and the MS. 

Fig.~\ref{Angle_diff_estimate} shows the average MSE of the angular difference estimate in the second stage. The increase of the time slots improves performance. This in turn brings higher RIS BF gain and average SE, as seen in Figs.~\ref{RIS_BF_gain} and~\ref{SE}. It is worth noting that for $T = 16$, the average SE of the proposed scheme is close to that with full CSI, even in the low SNR regime.



\section{Conclusion}
We studied the CE problem for the RIS-aided mmWave MIMO systems. We proposed a two-stage iterative reweighted method that finds estimates of the channel parameters in a sequential optimization loop. Simulation results confirmed the advantages of the proposed scheme compared to the two-stage OMP approach in terms of CE and SE performance. Since the used gradient descent method cannot guarantee the optimality of the solutions and its convergence may be slow, it is of interest to study more computational efficient algorithms in future work. 

\appendix
  \subsection{Derivatives in the First Stage}
  \label{FirstAppendix}
  
 The first-order partial derivative of $\tilde{S}(\hat{\bg}^{(i+1)},\theta, \phi)$ in \eqref{obj_func1} with respect to $\theta$ is written as 
\begin{equation}
    \frac{\partial \tilde{S}(\hat{\bg}^{(i+1)},\theta, \phi)}{\partial \theta}=  \sum_{t =1}^{{T}} - \frac{\partial \beta_t}{\partial \theta} z_t^* z_t - \beta_t \frac{\partial z_t^*}{\partial \theta} z_t- \beta_t z_t^* \frac{\partial z_t}{\partial \theta} ,\nonumber
\end{equation}
where $\frac{\partial \beta_t}{\partial \theta}$, $\frac{\partial z_t^*}{\partial \theta}$, $\frac{\partial z_t}{\partial \theta}$, and $\frac{\partial \|\bA_t\bX_t\|_{\mathrm{F}}^2}{\partial \theta}$ are given by 
\begin{align}
    \frac{\partial \beta_t}{\partial \theta} &= -\beta_t^2 \sum_{k =1}^{N_X}\bx_{t,k}^{\mathsf{H}} \frac{\partial \bA_t^{\mathsf{H}}}{\partial \theta}\bA_t\bx_{t,k} + \bx_{t,k}^{\mathsf{H}}  \bA_t^{\mathsf{H}} \frac{\partial \bA_t}{\partial \theta}\bx_{t,k}, \nonumber\\
    \frac{\partial z_t^*}{\partial \theta}& = \sum_{k =1}^{N_X}\by_{t,k}^{\mathsf{H}}  \frac{\partial \bA_t}{\partial \theta}\bx_{t,k}, \frac{\partial z_t}{\partial \theta} = \sum_{k =1}^{N_X}\bx_{t,k}  \frac{\partial \bA_t^{\mathsf{H}}}{\partial \theta}\by_{t,k}^{\mathsf{H}}, \nonumber\\
    \frac{\partial \|\bA_t\bX_t\|_{\mathrm{F}}^2}{\partial \theta} &=  \sum_{k =1}^{N_X}\bx_{t,k}^{\mathsf{H}} \frac{\partial \bA_t^{\mathsf{H}}}{\partial \theta}\bA_t\bx_{t,k} + \bx_{t,k}^{\mathsf{H}}  \bA_t^{\mathsf{H}} \frac{\partial \bA_t}{\partial \theta}\bA_t\bx_{t,k},\nonumber
\end{align}
and $\frac{\partial \bA_t}{\partial \theta} = \bA_t  \bD_{\theta}$, $\bD_{\theta}  = \mathrm{diag}\big([0, -j 2\pi \frac{d}{\lambda} \cos(\theta),\cdots,\newline -j 2\pi \frac{d}{\lambda} (i-1) \cos(\theta), \cdots, -j 2\pi \frac{d}{\lambda} (N_{\text{BS}}-1) \cos(\theta)]^{\mathsf{T}}\big)$.

Similarly, we can write the first-order partial derivative of $\tilde{S}(\hat{\bg}^{(i+1)},\theta, \phi)$ in \eqref{obj_func1} with respect to $\phi$ as
\begin{equation}
     \frac{\partial \tilde{S}(\hat{\bg}^{(i+1)},\theta, \phi)}{\partial \phi}=  \sum_{t =1}^T - \frac{\partial \beta_t}{\partial \phi} z_t^* z_t - \beta_t \frac{\partial z_t^*}{\partial \phi} z_t- \beta_t z_t^* \frac{\partial z_t}{\partial \phi} ,\nonumber
\end{equation}
and $\frac{\partial \bA_t}{\partial \phi}  = \bW_t^{\mathsf{H}} \bD_{\phi} \boldsymbol {\alpha}(\phi) \boldsymbol {\alpha}^{\mathsf{H}}(\theta)$, $\bD_{\phi}  = \mathrm{diag}\big([0, j 2\pi \frac{d}{\lambda}\cos(\phi),\newline\cdots, j 2\pi \frac{d}{\lambda} (i-1) \cos(\phi), \cdots, j 2\pi \frac{d}{\lambda} (N_{\text{MS}}-1) \cos(\phi)]^{\mathsf{T}}\big)$.

  \subsection{Derivatives in the Second Stage}
  \label{SecondAppendix}
    The derivative of $\tilde{J}(\hat{q}^{(i+1)}, \tilde{\theta})$ in \eqref{obj_second_step1} with respect to $\tilde{\theta}$ can be written as 
\begin{equation}
    \frac{\partial \tilde{J}(\hat{q}^{(i+1)}, \tilde{\theta})}{ \partial \tilde{\theta}} = - \frac{\partial \gamma}{\partial \tilde{\theta}}  \tau - \gamma \frac{\partial \tau}{\partial \tilde{\theta}},\nonumber
\end{equation}
where $\frac{\partial \gamma}{\partial \tilde{\theta}}$ and $\frac{\partial \tau}{\partial \tilde{\theta}}$ are in the form of 
\begin{align}
 \partial \gamma/ \partial \tilde{\theta} &= -\gamma^2\big(\boldsymbol {\alpha}(\tilde{\theta})^{\mathsf{H}} \bD_{\tilde{\theta}}^{\mathsf{H}}\tilde{\boldsymbol{\Omega}}^{\mathsf{H}}\tilde{\boldsymbol{\Omega}}\boldsymbol {\alpha}(\tilde{\theta}) + \boldsymbol {\alpha}(\tilde{\theta})^{\mathsf{H}} \tilde{\boldsymbol{\Omega}}^{\mathsf{H}}\tilde{\boldsymbol{\Omega}}\bD_{\tilde{\theta}}\boldsymbol {\alpha}(\tilde{\theta})\big),\nonumber\\
    \partial \tau / \partial \tilde{\theta} &=  \boldsymbol {\alpha}(\tilde{\theta})^{\mathsf{H}}\bD_{\tilde{\theta}}^{\mathsf{H}}\tilde{\boldsymbol{\Omega}}^{\mathsf{H}}\hat{\bg}\hat{\bg}^{\mathsf{H}} \tilde{\boldsymbol{\Omega}}\boldsymbol {\alpha}(\tilde{\theta}) +\boldsymbol {\alpha}(\tilde{\theta})^{\mathsf{H}}\tilde{\boldsymbol{\Omega}}^{\mathsf{H}}\hat{\bg}\hat{\bg}^{\mathsf{H}} \tilde{\boldsymbol{\Omega}}\bD_{\tilde{\theta}}\boldsymbol {\alpha}(\tilde{\theta}),\nonumber
\end{align}
and $\bD_{\tilde{\theta}}  = \mathrm{diag} \big([0, j 2\pi \frac{d}{\lambda} \cos(\tilde{\theta}),\cdots, j 2\pi \frac{d}{\lambda} (i-1) \cos(\tilde{\theta}), \cdots,\newline j 2\pi \frac{d}{\lambda} (N_{\text{RIS}}-1) \cos(\tilde{\theta})]^{\mathsf{T}}\big)$.

\bibliographystyle{IEEEtran}
\bibliography{IEEEabrv,Ref}

\end{document}